\documentclass[aip,graphicx]{revtex4-1}
\usepackage[T2A]{fontenc}
\usepackage[english]{babel}
\usepackage{amssymb,latexsym,amsmath,amscd}
\usepackage{graphicx,color,framed}
\begin{document}
\selectlanguage{english}

\title{The local phase transitions of the solvent in the neighborhood of a 
solvophobic polymer at high pressures}
\author{\firstname{Yu.~А.} \surname{Budkov}}
\email[]{urabudkov@rambler.ru}

\affiliation{G.A. Krestov Institute of Solution Chemistry of the Russian Academy of Sciences, Ivanovo, Russia}
\affiliation{National Research University Higher School of Economics, Moscow, Russia}

\author{\firstname{I.~I.} \surname{Vyalov}}
\affiliation{Istituto Italiano di Tecnologia, via Morego 30, Genova 16163, Italy}

\author{\firstname{ A.~L.} \surname{Kolesnikov}}
\affiliation{Ivanovo State University, Ivanovo, Russia}
\affiliation{Institut f\"{u}r Nichtklassische Chemie e.V., Universitat Leipzig,
Leipzig, Germany}
\author{\firstname{ N.} \surname{Georgi}}
\email[]{bancocker@mail.ru}
\affiliation{Max Planck Institute for Mathematics in the Sciences, Leipzig, Germany}

\author{\firstname{ G.~N.} \surname{Chuev}}
\affiliation{Max Planck Institute for the Physics of Complex Systems, Dresden, Germany}
\affiliation{Institute of Theoretical and Experimental Biophysics,
Russian Academy of Science, Pushchino, Moscow Region, Russia}

\author{\firstname{ M.~G.} \surname{Kiselev}}
\affiliation{G.A. Krestov Institute of Solution Chemistry of the Russian Academy of Sciences, Ivanovo, Russia}


\begin{abstract} 
We investigate local phase transitions of the solvent in the neighborhood 
of a solvophobic polymer chain which is induced by a change of the polymer-solvent 
repulsion and the solvent pressure in the bulk solution. We describe the polymer 
in solution by the Edwards model, where the conditional partition function 
of the polymer chain at a fixed radius of gyration is described by a mean-field theory. 
The contributions of the polymer-solvent and the solvent-solvent interactions to the total 
free energy are described within the mean-field approximation. 
We obtain the total free energy of the solution as a function of the radius of gyration 
and the average solvent number density within the gyration volume. 
The resulting system of coupled equations is solved varying the polymer-solvent repulsion strength 
at high solvent pressure in the bulk. We show that the coil-globule (globule-coil) transition occurs accompanied 
by a local solvent evaporation (condensation) within the gyration volume.
\end{abstract}
\maketitle

\section{Introduction}

In 2002 ten Wolde and David Chandler proposed a very elegant idea \cite{Chandler_2002} 
stating that a hydrophobic polymer chain immersed in an aqueous medium,
can undergo a coil-globule transition when in a neighborhood of the polymer a 
surface dewetting transition occurs.

The authors speculated, based on the results of computer simulations, that this effect
is reminiscent of the first-order phase transition. It should be noted that 
this statement amounts to proposing a fundamentally new mechanism of the polymer collapse, 
which is distinct from the standard mechanism adopted in statistical mechanics of macromolecules.
As is well known from the polymer statistical mechanics, when the solvent becomes poorer, 
the polymer coil shrinks leading eventually to a collapse of the 
polymer coil \cite{DeGennes,Grosberg_Khohlov}.

To describe the above mentioned mechanism within a theory it is therefore necessary 
to take the solvent explicitly into account. However, most  theoretical models  
describe the solvent only implicitly \cite{Edwards_1,Edwards_2,Birshtein_1,Birshtein_2,Moore,
Lifshitz,Lifshitz_Grosberg,deGennes_collaps,Muthukumar,Sanchez,Kholodenko,
Grosberg1,Grosberg2,Grosberg3,Grosberg4,Khohlov,Khohlov_2}, 
i.e. its influence on the macromolecule is taken into account through 
an effective monomer-monomer interaction. Such an approach simplifies the model, 
however the details of the solvent behavior --  in particular phase transitions in the bulk solution, 
heterogeneity of the solvent near the polymer chain, and the dependence 
of the solvent quality on the pressure -- are not taken into account. 
At present only few examples exist considering the solvent explicitly.

In the pioneering works of Flory and Schultz \cite{Flory_Schultz} (mean-field theory), 
de Gennes and Brochard \cite{deGennes2} (scaling approach) and the works of Vilgis 
and co-authors \cite{Vilgis,Vilgis_Dua,Dua} (path integral methodology) 
describing the conformational behavior of a polymer chain in a vicinity of the 
binary mixture critical point the solvent has been taken into account explicitly. 
The authors in \cite{Lifshitz_Grosberg}  briefly discussed  the influence of the 
solvent on the spatial structure of the globule. In \cite{Erukhimovich_JETP} 
a field-theoretical approach has been adopted to investigate the density 
of the globule in a critical solvent.

In the work of Matsuyama and Tanaka \cite{Tanaka} the conformational 
phase transition of an isolated polymer chain has been approached. 
The chain was allowed to form  physical bonds with explicitly present solvent molecules.
On the basis of a Flory mean-field type theory, a formula for the temperature dependence 
of the expansion factor of the chain has been derived. The formation 
of  physical bonds between the polymer and the solvent 
molecules was shown to cause a re-entrant conformational change between a coiled and 
a globular state when the temperature was varied. Tanaka et al \cite{Tanaka2} considered 
a model for poly(N-isopropylacrylamide) chain in a water - 
methanol mixture on the basis of  statistical mechanics.  
The model included a competitive water-polymer and methanol-polymer hydrogen bond formation. 
The obtained mean squared end-to-end distance has been compared with experiments.
In the work \cite{Tamm} have classified the various types of the 
coil-globule transition of a polymer chain in an associating solvent taking 
into account the microscopic nature of such the solvent.

Unfortunately,  most theoretical models of dilute polymer solutions with explicit 
account of the solvent that were mentioned above cannot describe the heterogeneity of the solvent in the neighborhood 
of the polymer chain depending on its conformation. This is mainly related to the fact 
that polymer solutions are treated as incompressible \cite{Tanaka,Tanaka2,Tamm}, which is incorrect 
in the case of strong polymer-solvent repulsion. Indeed, in the case of strong polymer-solvent repulsion 
in a vicinity of the polymer chain cavities can form  which is not possible 
within the approximation of incompressible solution. Thus in this case one can expect the appearance 
of a liquid-gas transition in a neighborhood of a polymer induced by the strong polymer-solvent repulsion. 
Therefore, a theory describing the conformational transition of the polymer triggered by the solvent, 
requires an explicit account of the solvent to capture its inhomogeneity close to the polymer backbone.

In the present work such a self-consistent field model is developed. The study presented here 
is based on the formalism, developed in our previous work \cite{Budkov_2014}. 
In contrast to our previous investigations where the solvent-solvent interactions were
purely repulsive in the present work we lift this restriction by describing the low-molecular 
weight solvent via a Van-der-Waals equation of state. This allows to study the solvophobic polymer 
chain in a wide range of temperature and solvent density. We consider the theory beyond the
approximation of incompressible solution so that volume fractions of monomers and solvent 
molecules within globule volume are considered as independent variables. We investigate 
the regime of strong repulsion between monomers and solvent molecules. As the polymer-solvent repulsion 
strength increases the collapse to a globular state occurs accompanied by a local solvent evaporation in the 
neighborhood of the macromolecule. However, if the pressure of the low-molecular weight solvent 
(at fixed polymer-solvent repulsion strength) in the bulk solution exceeds a threshold value, 
the polymer expands from globular to a coiled configuration accompanied by a condensation 
of the solvent near the polymer backbone.

The paper is organized as follows. In Sec. II, we present our theoretical formalism, in Sec. III 
the limiting regimes are analyzed, in section IV we provide numerical results and 
their discussion and in Sec. V we summarize our findings.

\section{Theory}
We consider a polymer chain molecule immersed in a low-molecular weight solvent 
at a specified number density. As already mentioned in the introductory section, 
in contrast to our previous investigations \cite{Budkov_2014} 
(solvent-solvent interaction is purely repulsive) we consider the solvent 
to obey the Van-der-Waals equation of state. Moreover, we shall describe the 
polymer-solvent interactions as purely repulsive. In other words, we assume that 
the polymer chain is solvophobic with respect to the solvent.
We would like to stress that throughout this paper the term "solvophobic" denotes 
the effective repulsive polymer-solvent interaction. Our goal is to study 
the dependence of the polymer chain conformation and the behavior of 
the solvent near the polymer chain as a function of solvent pressure and the monomer-solvent
repulsion strength. We shall consider the polymer in the framework of the Edwards model 
\cite{Edwards_1,Edwards_2}.

In this work we will employ a simple formalism reminiscent of the classical Flory type theories describing 
the behavior of the polymer chain in terms of the expansion factor or the radius of gyration. 
A more rigorous theory for description of the coil-globule transitions 
has been developed in works of Lifshitz and co-authors \cite{Lifshitz,Lifshitz_Grosberg} 
based on the idea that the globule can be treated as a fragment of the semi-dilute polymer solution. 
In contrast to Flory type theories the behavior of a globule 
within the Lifshitz theory has been described in terms of density functional theory, so that such phenomena 
as surface tension and fluctuation of globule's surface were taken into account \cite{Lifshitz_Grosberg}. 
In the present work we use a Flory type theory since it is simpler and widely used.

We start from the conditional partition function of the solution, which can be written as follows
\begin{equation}
\label{Z}
Z(R_{g})=\int d\Gamma_{p}(R_{g})\int d\Gamma_{c}\exp\left[-\beta H_{p}-\beta H_{s}-\beta H_{ps}\right],
\end{equation}
where the symbol $\int d\Gamma(R_{g})(..)$ denotes the integration over microstates
of the polymer chain at a fixed radius of gyration $R_{g}$; the symbol 
$\int d\Gamma_{s}(..)=\frac{1}{N_{s}!}\int\limits_{V} d\bold{r}_{1}..\int\limits_{V} d\bold{r}_{N_{s}}(..)$
denotes the integration over solvent molecules coordinates; $N_{s}$ is the total number 
of solvent molecules, $V$ is a volume of the system;
\begin{equation}
\beta H_{p}=\frac{w_{p}}{2}\int\limits_{0}^{N}ds_{1}\int\limits_{0}^{N}ds_{2}\delta\left(\bold{r}(s_{1})-\bold{r}(s_{2})\right)=
\frac{w_{p}}{2}\int d\bold{x}\hat{\rho}_{p}^2(\bold{x})
\end{equation}
is the Hamiltonian of the monomer-monomer excluded volume interaction;
$w_{p}$ is the second virial coefficient of the monomer-monomer excluded volume interaction and
$\hat{\rho}_{p}(\bold{x})=\int\limits_{0}^{N} ds\delta(\bold{x}-\bold{r}(s))$
is the monomer microscopic density; $N$ is degree of polymerization of the polymer chain;
\begin{equation}
\beta H_{ps}=w_{ps}\int\limits_{0}^{N}ds\sum\limits_{j=1}^{N_{s}}\delta\left(\bold{r}(s)-\bold{r}_{j}\right)
=w_{ps}\int d\bold{x}\hat{\rho}_{p}(\bold{x})\hat{\rho}_{s}(\bold{x})
\end{equation}
is the Hamiltonian of the polymer-solvent interaction;
$w_{ps}$ is the second virial coefficient for the polymer-solvent interaction 
(we call it the solvophobic strength) and
$\hat{\rho}_{s}(\bold{x})=\sum\limits_{i=1}^{N_{s}}\delta\left(\bold{x}-\bold{r}_{i}\right)$
is the microscopic density of the solvent molecules;
\begin{equation}
H_{s}=\frac{1}{2}\sum\limits_{j\neq{i}}V(\bold{r}_{i}-\bold{r}_{j})=\frac{1}{2}\sum\limits_{j\neq{i}}(V_{hc}(\bold{r}_{i}-\bold{r}_{j})+
V_{att}(\bold{r}_{i}-\bold{r}_{j}))
\end{equation}
is the Hamiltonian of solvent-solvent interaction;
\begin{equation}
V_{hc}(\bold{r})=\Biggl\{
\begin{aligned}
\infty,\quad&|\bold{r}|\leq d_{s}\,\\
0,\quad& |\bold{r}|> d_{s}
\end{aligned}
\end{equation}
is the hard-core potential ($d_{s}$ is the solvent molecule diameter); 
$V_{att}(\bold{r})=-a_{s}\delta(\bold{r})$ ($a_{s}>0$) is the attractive part 
of the total potential of the solvent-solvent interaction.

The conditional partition function of the polymer solution at a fixed radius
of gyration $R_{g}$ of the polymer chain at the level of the mean-field approximation 
is derived from  the partition function of the solution, which takes the form
\begin{eqnarray}
Z(R_{g})=\int d\Gamma_{p}(R_{g})e^{-\beta H_{p}}\int d\Gamma_{c}e^{-\beta H_{s}-\beta H_{ps}}=
Z_{p}(R_{g})\int d\Gamma_{s}e^{-\beta H_{s}}\left<e^{-\beta H_{ps}}\right>_{p},
\end{eqnarray}
where
\begin{equation}
Z_{p}(R_{g})=\int d\Gamma (R_{g})e^{-\beta H_{p}}
\end{equation}
is the polymer partition function; 
the symbol $\left<(..)\right>_{p}=\frac{1}{Z_{p}(R_{g})}\int d\Gamma(R_{g})e^{-\beta H_{p}}(..)$
denotes  averaging over polymer microstates with a fixed radius of gyration.
Using cumulant expansion \cite{Kubo} and truncating at the first order we obtain
\begin{equation}
\left<e^{-\beta H_{ps}}\right>_{p}\approx e^{-\beta\left<H_{ps}\right>_{p}}.
\end{equation}
Therefore 
\begin{equation}
\beta\left<H_{ps}\right>_{p}=w_{ps}\int\limits_{V} d\bold{x}\hat{\rho}_{s}(\bold{x})\left<\hat{\rho}_{p}(\bold{x})\right>_{p}\simeq \frac{Nw_{ps}}{V_{g}}
\int\limits_{V_{g}}d\bold{x}\hat{\rho}_{s}(\bold{x}),
\end{equation}
where the approximation
\begin{equation}
 \left<\hat {\rho}_{p}(\bold{x})\right>_{p}\simeq \Biggl\{
\begin{aligned}
\frac{N}{V_{g}},\quad&|\bold{x}|\leq R_{g}\,\\
0,\quad& |\bold{x}|> R_{g}
\end{aligned}
\end{equation}
has been introduced; $V_{g}=\frac{4\pi  R_{g}^3}{3}$ is a value of the gyration volume.
This results in the following expression for the partition function of the solution
\begin{equation}
Z(R_{g})=Z_{p}(R_{g})Z_{s}(R_{g}),
\end{equation}
where $Z_{s}(R_{g})$ has the form
\begin{eqnarray}
Z_{s}(R_{g})=\int d\Gamma_{s}e^{-\beta H_{s}-\frac{w_{ps}N}{V_{g}}\int\limits_{V_{g}}d\bold{x}\hat{\rho}_{s}(\bold{x})}=
\frac{1}{N_{s}!}\int\limits_{V} d\bold{r}_{1}..\int\limits_{V} d\bold{r}_{N_{s}}
e^{-\beta H_{s}-\frac{w_{ps}N}{V_{g}}\int\limits_{V_{g}}d\bold{x}\hat{\rho}_{s}(\bold{x})}.
\end{eqnarray}
The last expression can be written as a sum
\begin{equation}
Z_{s}(R_{g})=\sum\limits_{n=0}^{N_{s}}\mathcal{Z}_{s}(R_{g},n),
\end{equation}
where
\begin{eqnarray}
&&\mathcal{Z}_{s}(R_{g},n)=\frac{e^{-\frac{w_{ps}Nn}{V_{g}}}}{(N_{s}-n)!n!}
\int\limits_{V_{g}}d\bold{x}_{1}..\int\limits_{V_{g}}d\bold{x}_{n}\int\limits_{V-V_{g}}d\bold{y}_{1}..\int\limits_{V-V_{g}}d\bold{y}_{N_{s}-n}
e^{-\beta H_{s}}
\end{eqnarray}
is the solvent partition function with $n$ being the number of solvent molecules in the gyration volume. 
In order to evaluate $Z_s(Rg, n)$ we introduce the approximation $H_s=H_s^{n}+ H_s^{N_s-n}$ ($H_s=H_s^{N_{s}}$) 
which is accurate for sufficiently large gyration volumes, so that all interface effects may be safely neglected. 
In other words, the surface layer does not contribute to the total free energy.  
Applying the above mentioned assumption and the mean-field approximation, we obtain
\begin{eqnarray}
\label{Zc}
Z_{s}(R_{g})\simeq\sum\limits_{n=0}^{N_{s}}\mathcal{Z}_{s}^{(b)}(R_{g},n)\mathcal{Z}_{s}^{(g)}(R_{g},n),
\end{eqnarray}
where 
\begin{equation}
\mathcal{Z}_{s}^{(b)}(R_{g},n)=\frac{(V-V_{g})^{N_{s}-n}}{(N_{s}-n)!}e^{-\beta F_{ex,s}^{(b)}(R_{g},n)},
\end{equation}
\begin{equation}
\mathcal{Z}_{s}^{(g)}(R_{g},n)=\frac{V_{g}^n}{n!}e^{-\beta F_{ex,s}^{(g)}(R_{g},n)},
\end{equation}
\begin{equation}
\label{F_ex1}
\beta F_{ex,s}^{(b)}=-(N_{s}-n)\ln\left(1-\frac{(N_{s}-n)v_{s}}{V-V_{g}}\right)-\frac{\beta a_{s}(N_{s}-n)^2}{V-V_{g}},
\end{equation}
\begin{equation}
\label{F_ex2}
\beta F_{ex,s}^{(g)}=\frac{w_{ps}Nn}{V_{g}}-n\ln\left(1-\frac{nv_{s}}{V_{g}}\right)-\frac{\beta a_{s}n^2}{V_{g}},
\end{equation}
$a_{s}$ is a Van-der-Waals attraction parameter for the solvent, and
$v_{s}$ is an excluded volume of solvent molecules.

In the sum (\ref{Zc}) only the highest order term is non-negligible. This is related to the fact that 
the deviation of the number of molecules of a liquid in some sufficiently large 
volume is very small compared to the average number, due to very small liquid compressibility. 
This term corresponds to the number $n=N_{1}$ which can be obtained from the extremum condition
\begin{eqnarray}
\frac{\partial}{\partial{n}}\ln\mathcal{Z}_{s}(R_{g},n)=0.
\end{eqnarray}
Therefore we arrive at the expression:
\begin{eqnarray} 
\mathcal{Z}_{s}(R_{g},N_{1})=\mathcal{Z}_{s}^{(b)}(R_{g},N_{1})\mathcal{Z}_{s}^{(g)}(R_{g},N_{1}).
\end{eqnarray}
Therefore we assume that the volume of the system consists of two parts: the gyration volume containing
predominantly monomers of the polymer chain and a bulk solution. We consider the solvent 
concentration at equilibrium in the two subvolumes varying the strength
of interaction of the polymer-solvent. The partition function of the solvent 
within mean-field approximation can then be written as the product:
\begin{equation} \label{PartSum:Total}
Z(R_{g})=\mathcal{Z}_{s}^{(b)}(R_{g},N_{1})\mathcal{Z}_{s}^{(g)}(R_{g},N_{1})Z_{p}(R_{g}).
\end{equation}
The number of the solvent molecules in the gyration volume $N_{1}$ satisfies the extremum condition
\begin{equation}
\frac{\partial}{\partial{N_{1}}}\ln{\mathcal{Z}_{s}^{(b)}(R_{g},N_{1})\mathcal{Z}_{s}^{(g)}(R_{g},N_{1})}=0.
\end{equation}

The equilibrium value of the radius of gyration is determined from the minimum of the total 
free energy of the solution. The conditional free energy of the polymer 
chain takes the form: 
\begin{equation}
\label{FreeEnerTotal}
\beta F_{p}(R_{g})=-k_{B}T\ln{Z}_{p}(R_{g})=\beta F_{id,p}(R_{g})+\beta F_{ex,p}(R_{g}),
\end{equation}
where $F_{id,p}(R_{g})$ is a conditional free energy of the ideal Gaussian polymer chain which can be
determined by the following interpolation formula \cite{Fixman_Gyration,Birshtein_1,Birshtein_2}:
\begin{eqnarray}
\label{FreeEnerInterIdeal}
\beta F_{id,p}(\alpha)=\frac{9}{4}\left(\alpha^2+\frac{1}{\alpha^2}\right)
\end{eqnarray}
where $\alpha=R_{g}/R_{0g}$ denotes the expansion factor, $R_{0g}^2=Nb^2/6$
is the mean-square radius of gyration of the ideal polymer chain and $b$ is the Kuhn length of the segment, 
$\beta =1/k_{B}T$ is an inverse temperature, $k_{B}$ is a Boltzmann constant.
The second term in (\ref{FreeEnerTotal}) is an excess conditional free energy which can be 
written within mean-field approximation as:
\begin{equation}
\label{FreeEnerInterMF}
\beta F_{ex,p}=\frac{w_{p}N^2}{2V_{g}}=\frac{9\sqrt{6}w_{p}\sqrt{N}}{4\pi b^{3}\alpha^3}.
\end{equation}
Using the expressions (\ref{FreeEnerInterIdeal}) and (\ref{FreeEnerInterMF}), we obtain 
the following expression for conditional free energy of the polymer chain
\begin{eqnarray}
\label{FreeEnerInterFinal}
\beta F_{p}(\alpha)=\frac{9}{4}\left(\alpha^2+\frac{1}{\alpha^2}\right)+\frac{9\sqrt{6}\tilde{w}_{p}\sqrt{N}}{4\pi\alpha^3},
\end{eqnarray}
where the dimensionless second virial coefficient of the monomer-monomer interaction 
$\tilde{w}_{p}=w_{p}b^{-3}$ has been introduced. The first and second terms in (\ref{FreeEnerInterFinal}) 
are determined by the conformational entropy of an ideal Gaussian polymer. 
The third term determines the contribution to the polymer free energy of 
the  monomer-monomer volume interaction at the level of second order virial expansion. 
We shall show below that such approximation for the volume interaction contribution 
is sufficient to describe the coil-globule transition due to the solvent effect. This is in contrast 
to  the implicit solvent treatment where the polymer free energy is commonly expanded up 
to the third order in the monomer-monomer volume interactions 
\cite{Grosberg_Khohlov,Grosberg1,Grosberg2,Grosberg3,Grosberg4}. 
In present study for simplicity we neglect a contribution of the monomer concentration 
fluctuations within the gyration volume into the polymer free energy. 
The latter is motivated by the fact that fluctuations of monomers concentration 
as one can show lead to small correction into the polymer free energy 
and consequently can be omitted.

The expression for the solvent Helmholtz free energy takes the form
\begin{eqnarray} 
\label{FreeEnersolvent1}
F_{s}(R_{g},N_{1})= F_{s}^{(b)}(R_{g},N_{1})+ F_{s}^{(g)}(R_{g},N_{1}),
\end{eqnarray}
where $\beta F_{s}^{(b)}(R_{g},N_{1})=-\ln{\mathcal{Z}_{s}^{(b)}(R_{g},N_{1})}$, 
and $\beta F_{s}^{(g)}(R_{g},N_{1})=-\ln{\mathcal{Z}_{s}^{(g)}(R_{g},N_{1})}$.
Minimizing $\beta F_{s}(R_{g},N_{1})$ with respect to $N_{1}$ and 
introducing the dimensionless solvent number densities $\tilde{\rho}_{1}=N_{1}b^3/V_{g}$ 
and $\tilde{\rho}=N_{s}b^3/V$, temperature $\tilde{T}=k_{B}T b^3/a_{s}$, 
and excess chemical potential of the solvent $\tilde{\mu}_{ex,s}=\mu_{ex,s}b^3/a_{s}$ 
we finally obtain the equation for the concentration of the solvent $\tilde{\rho}_{1}$ within the gyration volume
\begin{eqnarray}  
\label{rho1}
\tilde{\rho}_{1}= \tilde{\rho}\exp\left[-\frac{9\sqrt{6}\tilde{w}_{ps}}{2\pi \sqrt{N}\alpha^3}+
\frac{\tilde{\mu}_{ex,s}(\tilde{\rho},\tilde{T})-\tilde{\mu}_{ex,s}(\tilde{\rho}_{1},\tilde{T})}{\tilde{T}}\right],
\end{eqnarray}
which is valid for $V\gg V_{g}$ and $N_{s}\gg N_{1}$. It should be noted that expression (\ref{rho1})
provides a condition for the equality of solvent chemical potentials in the gyration volume and in the bulk solution.
The excess chemical potential of the solvent within the mean-field approximation has a form
\begin{equation}
\label{muexc}
\tilde{\mu}_{ex,s}(\tilde{\rho},\tilde{T})=
\frac{\tilde{T}\tilde{\rho} \tilde{v}_{s}}{1-\tilde{\rho} \tilde{v}_{s}}-
\tilde{T}\ln\left(1-\tilde{\rho} \tilde{v}_{s}\right)-2\tilde{\rho},
\end{equation}
where $\tilde{v}_{s}=v_{s}b^{-3}$ is a dimensionless excluded volume of the solvent molecules.
We would like to emphasize that expression for the solvent excess chemical potential (\ref{muexc}) 
presupposes a gas-liquid transition in the bulk solution, therefore such phase transition 
can be realized within gyration volume due to polymer-solvent interaction.

Further, using the equations (\ref{FreeEnerInterFinal}-\ref{rho1}), 
and  taking the derivative of the total free energy with respect to $\alpha$ 
and equating it to zero, after some algebra we obtain
\begin{eqnarray} \label{AlphaPolynom}
\alpha^5-\alpha=\frac{3\sqrt{6}}{2\pi}\tilde{w_{p}}\sqrt{N}+\frac{2}{3}N\tilde{w}_{ps}\tilde{\rho}_{1}\alpha^3-
\frac{2\pi\sqrt{6}}{81}N^{3/2}\alpha^6\frac{\tilde{P}(\tilde{\rho},\tilde{T})-\tilde{P}(\tilde{\rho_{1}},\tilde{T})}{\tilde{T}},
\end{eqnarray}
where $\tilde{w}_{ps}=w_{ps}b^{-3}$.
In addition, in (\ref{AlphaPolynom}) we have introduced the dimensionless
solvent pressure $\tilde{P}=Pb^6/a_{s}=\tilde{\rho}\frac{\partial\tilde{f}_{s}}{\partial\tilde{\rho}}-\tilde{f}_{s}$ 
($f_{s}=F_{s}/V$ is the density of the solvent Helmholtz free energy; $\tilde{f_{s}}=f_{s}b^6/a_{s}$ 
is a dimensionless density of the solvent free energy), 
which within our model satisfies the  well-known Van-der-Waals equation of state
\begin{equation}
\label{Pressure}
\tilde{P}(\tilde{\rho},\tilde{T})=\frac{\tilde{\rho}\tilde{T}}{1-\tilde{\rho}\tilde{v}_{s}}-\tilde{\rho}^2.
\end{equation}

The first term on the right hand side of the equation (\ref{AlphaPolynom}) 
is related to the monomer-monomer excluded volume interaction. 
The second  term is related to the polymer-solvent interaction. 
The third term is proportional to the difference between the solvent 
pressure within gyration volume and the bulk.

\section{Analysis of limiting regimes}

In this section we analyze the limiting regimes for the 
radius of gyration, which follows from equations (\ref{rho1}) and (\ref{AlphaPolynom}).

In case of $\tilde{w}_{ps}\ll 1$ a swelling regime occurs $\alpha \sim \tilde{w}_{p}^{1/5}N^{1/10}$
$\left(\frac{R_{g}}{b}\sim \tilde{w}_{p}^{1/5}N^{3/5}\right)$ which is well known
from the classical Flory mean-field theory \cite{Flory_book}.

Considering the opposite regime - $\tilde{w}_{ps}\gg 1$, i.e. when the solvent-polymer 
interaction is strongly repulsive and $\tilde{\rho}_{1}\ll \tilde{\rho}$, the equation (\ref{AlphaPolynom}) simplifies to
\begin{eqnarray} 
\label{AlphaPolynomSimple}
\alpha^5-\alpha=
\frac{3\sqrt{6}}{2\pi}\tilde{w_{p}}\sqrt{N}+\frac{2}{3}N\tilde{w}_{ps}\tilde{\rho}_{1}\alpha^3-
\frac{2\pi\sqrt{6}}{81}N^{3/2}\frac{\tilde{P}(\tilde{\rho},\tilde{T})}{\tilde{T}}\alpha^6.
\end{eqnarray}
If the second term on the right hand side of the equation (\ref{AlphaPolynomSimple}) 
dominates then neglecting all except the second and  third terms on the right hand side we obtain 
a simple limiting law for the expansion factor $\alpha$ and the radius of gyration
\begin{eqnarray}
\label{AlphaRg2}
&&\alpha \simeq \left(\frac{9\sqrt{6}}{2\pi}\right)^{1/3}\left(\frac{\tilde{w}_{ps}\tilde{\rho}_{1}\tilde{T}}{\tilde{P}}\right)^{1/3}N^{-\frac{1}{6}},~~
\frac{R_{g}}{b}\simeq \frac{\sqrt{6}}{6}\left(\frac{9\sqrt{6}}{2\pi}\right)^{1/3} \left(\frac{\tilde{w}_{ps}\tilde{\rho}_{1}\tilde{T}}{\tilde{P}}\right)^{1/3}N^{1/3},
\end{eqnarray}
which provide the estimate of the size of the globule.
In this regime, the size of the globule is determined by a competition 
between the polymer-solvent repulsion which tends to expand the polymer chain
and the solvent pressure effect which tends to shrink it. 
Using the relations (\ref{AlphaRg2}) the monomer number density 
$\tilde{\rho}_{g}=Nb^3/V_{g}$ within the globule volume can be expressed 
as a function of the solvent pressure in the bulk and the temperature:
\begin{equation}
\label{globule_conc2}
\tilde{\rho}_{g}\sim \frac{\tilde{P}}{\tilde{w}_{ps}\tilde{\rho}_{1}\tilde{T}}.
\end{equation} 
If the first term on the right hand side of the equation (\ref{AlphaPolynomSimple}) 
dominates an analogous simple limiting law for the expansion factor $\alpha$ and the radius of gyration is obtained
\begin{eqnarray} 
\label{AlphaRg}
&&\alpha \simeq \left(\frac{243}{4\pi^2}\right)^{1/6}\left(\frac{\tilde{w_{p}}\tilde{T}}{\tilde{P}}\right)^{1/6}N^{-\frac{1}{6}},~~
\frac{R_{g}}{b}\simeq \frac{\sqrt{6}}{6}\left(\frac{243}{4\pi^2}\right)^{1/6} \left(\frac{\tilde{w_{p}}\tilde{T}}{\tilde{P}}\right)^{1/6}N^{1/3},
\end{eqnarray}
which corresponds to a globular conformation. In this case, the size of the globule is determined
by a competition between the solvent pressure which tends to shrink the polymer chain
and the monomer excluded volume effect which tends to expand it. The monomer number density 
within the globule volume as function of solvent pressure in 
the bulk and temperature is then given by:
\begin{equation}
\label{globule_conc}
\tilde{\rho}_{g}\sim \left(\frac{\tilde{P}}{\tilde{w}_{p}\tilde{T}}\right)^{1/2}.
\end{equation}

We would like to stress that for both globular regimes (\ref{AlphaRg2}) and (\ref{AlphaRg})
following from the equation (\ref{muexc}) the solvent number density within gyration volume 
$\tilde{\rho}_{1}$ is independent of the degree of polymerization $N$.

In the regime of a dense solvent in the bulk, when $\tilde{\rho}\sim 1/\tilde{v}_{s}$ 
the bulk pressure exceeds the polymer-solvent repulsion leading to an expansion of the polymer chain. 
In this case $\tilde{\rho}_{1}=\tilde{\rho}+\delta \tilde{\rho}$, 
where $\delta\tilde{\rho}\ll \tilde{\rho}$. Thus we can 
expand the functions $\tilde{P}(\tilde{\rho}_{1},\tilde{T})$ 
and $\tilde{\mu}_{ex,s}(\tilde{\rho}_{1},\tilde{T})$ in a power series 
with respect to $\delta\tilde{\rho}$. Truncating the power series at 
the first term we obtain
\begin{equation}
\label{expan1}
\tilde{P}(\tilde{\rho},\tilde{T})-\tilde{P}(\tilde{\rho}_{1},\tilde{T})=-\frac{\partial{\tilde{P}}(\tilde{\rho},\tilde{T})}{\partial{\tilde{\rho}}}\delta\tilde{\rho},
\end{equation}
and
\begin{equation}
\label{expan2}
\tilde{\mu}_{ex,s}(\tilde{\rho},\tilde{T})-\tilde{\mu}_{ex,s}(\tilde{\rho}_{1},\tilde{T})=-
\frac{\partial{\tilde{\mu}_{ex,s}}(\tilde{\rho},\tilde{T})}{\partial{\tilde{\rho}}}\delta\tilde{\rho}.
\end{equation}
Thus, at first order in $\delta\tilde{\rho}$ the equation (\ref{rho1}) simplifies to
\begin{equation}
\label{delta}
\delta\tilde{\rho}=-\frac{9\sqrt{6}\tilde{w}_{ps}}{2\pi\sqrt{N}\alpha^3}\frac{\tilde{\rho}}{1+\frac{\tilde{\rho}}{\tilde{T}}
\frac{\partial{\tilde{\mu}_{ex,s}}(\tilde{\rho},\tilde{T})}{\partial{\tilde{\rho}}}}.
\end{equation}
Further, using an identity
\begin{equation}
\frac{1}{\tilde{T}}\frac{\partial{\tilde{P}}(\tilde{\rho},\tilde{T})}{\partial{\tilde{\rho}}}=1+\frac{\tilde{\rho}}
{\tilde{T}}\frac{\partial{\tilde{\mu}_{ex,s}}(\tilde{\rho},\tilde{T})}{\partial{\tilde{\rho}}},
\end{equation} 
and relations (\ref{expan1}-\ref{delta}) we arrive at the following equation with respect to $\alpha$
\begin{equation}
\alpha^5-\alpha=\frac{3\sqrt{6}}{2\pi}\left(\tilde{w_{p}}-\tilde{w}_{ps}^2\tilde{\rho}^2\tilde{T}\tilde{\chi}_{T}(\tilde{\rho},\tilde{T})\right)\sqrt{N},
\end{equation}
where we have introduced an isothermal compressibility
$\tilde{\chi}_{T}(\tilde{\rho},\tilde{T})=\frac{1}{\tilde{\rho}}\left(\frac{\partial{\tilde{\rho}}}{\partial{\tilde{P}}}\right)_{\tilde{T}}$ 
which within our model is determined by an expression
\begin{equation}
\tilde{\chi}_{T}(\tilde{\rho},\tilde{T})=
\frac{(1-\tilde{\rho}\tilde{v}_{s})^2}{\tilde{\rho}
\tilde{T}\left(1-\frac{2\tilde{\rho}}{\tilde{T}}(1-\tilde{\rho}\tilde{v}_{s})^2\right)}.
\end{equation}
Therefore, as a result we obtain that at large solvent densities in the bulk the 
monomer-monomer interactions are renormalized
\begin{equation}
\label{renorm}
\tilde{w}=\tilde{w}_{p}-\tilde{w}_{ps}^2\tilde{\rho}^2\tilde{T}\tilde{\chi}_{T},
\end{equation}
so that the expansion factor and the radius of gyration are determined by the following relations 
\begin{equation}
\label{coil}
\alpha \sim \tilde{w}^{1/5}N^{1/10}, ~~~ \frac{R_{g}}{b}\sim \tilde{w}^{1/5}N^{3/5}.
\end{equation}

The limiting behavior (\ref{renorm}) is a well known result, which was first 
obtained by Flory and Schultz \cite{Flory_Schultz} and de Gennes and Brochard \cite{deGennes2} 
for a homopolymer dissolved in a binary mixture in the vicinity of the critical point using 
a mean-field theory and scaling approach, respectively. This result has also been obtained 
within a field-theoretical approach by Vilgis et al \cite{Vilgis} and Erukhimovich \cite{Erukhimovich_JETP} 
in the framework of the Gaussian approximation. As is well known, the Gaussian approximation 
is adequate for simple non ionic liquids at large concentrations only \cite{Hansen}. 
Here the result (\ref{renorm}) has been obtained as a limiting case at 
the large solvent density.

\section{Numerical results and discussion}
Turning to the numerical analysis of the system of equations (\ref{rho1}) and (\ref{AlphaPolynom})
we fix the Van-der-Waals volume of the solvent molecule $\tilde{v}_{s}=1$, the  monomer-monomer 
volume interaction parameter $\tilde{w}_{p}=1$, and the degree of polymerization to $N=10^{3}$.

We first discuss the case when the temperature $\tilde{T}$ of the system is
below the critical temperature $\tilde{T}_{c}=\frac{8}{27}$ of the solvent ($\tilde{T}<\tilde{T}_{c}$) 
for different solvophobic strength $\tilde{w}_{ps}$. Hence, we consider the isotherm 
$\tilde{P}=\tilde{P}(\tilde{\rho},\tilde{T})$ increasing $\tilde{\rho}$ starting from the binodal.
Fig. 1 (a) shows the solvent number density in the gyration volume as a function of the solvent pressure 
in the bulk at a fixed solvophobic strength $\tilde{w}_{ps}$. Increasing the pressure $\tilde{P}$ in the bulk, 
the solvent concentration in the gyration volume decreases monotonically and at some threshold value jumps to a value
which is very close to the bulk solvent number density. The expansion factor in this range 
(Fig.1 (b)) abruptly changes the globular regime (\ref{AlphaRg}) and then jumps 
to the regime of the polymer coil (\ref{coil}). The jumps of $\tilde{\rho}_{1}$ and $\alpha$ 
are a consequence of the penetration of the solvent into the gyration volume leading to the 
equilibration of pressures between the gyration volume and the bulk solution. 
This amounts to a gas-liquid transition of the solvent within the gyration volume (Fig. 1 (b)).

Now we turn to the discussion of the region where $\tilde{T}>\tilde{T}_{c}$. 
We increase the solvent number density from critical one (Fig. 2 (a,b)), 
so that we consider behavior of polymer chain in the supercritical region of solvent. 
The results for a polymer chain in the supercritical solvent region  
($\tilde{T}>\tilde{T}_{c}$, $\tilde{\rho}>\tilde{\rho}_{c}$), 
increasing the solvent number density up from critical point are shown in (Fig. 2 (a,b)).

Quite similar behavior occurs in the present case compared to the region where 
$\tilde{T}<\tilde{T}_{c}$. The presence of jumps in the solvent concentration $\tilde{\rho}_{1}$ and
in the expansion factor $\alpha$ are also here due to the effect of the solvent
molecules intruding into the gyration volume which in turn leads to an
equilibrium between the pressure in the gyration volume and the bulk.

It is instructive to regard the dependence of the expansion factor 
on the solvophobic strength in the regions below ($\tilde{T}>\tilde{T}_{c}$) 
and above ($\tilde{T}<\tilde{T}_{c}$) the critical isotherm of the solvent. 
Fig.3 (a) shows such dependence at fixed bulk solvent concentration. 
In both cases a polymer chain collapse occurs to the regime (\ref{AlphaRg}) 
when the solvophobic strength exceeds a threshold value. It should be noted, that the 
polymer chain collapse in region $\tilde{T}>\tilde{T}_{c}$ occurs
at higher solvophobic strength than in the region $\tilde{T}<\tilde{T}_{c}$.
We would also like to stress that the polymer chain collapse occurs
as a first-order phase transition, confirming the hypothesis of 
ten Wolde and Chandler \cite{Chandler_2002}. Indeed, as shown in Fig. 3 (b) 
the solvent concentration in the gyration volume attains to very small 
values when the polymer chain collapse takes place.
In the region below the solvent critical isotherm this jump corresponds 
to a dewetting transition which results in a formation of a polymer globule surface surrounded 
by a layer of solvent gas. In the region above the critical isotherm a similar mechanism 
is responsible for the transition. However, in this case the collapse is caused 
by a layer of a gas-like fluid.

It should be noted that at higher degree of polymerization $N$ 
the globule expansion is more pronounced. The Fig.4 shows the expansion 
factor $\alpha$ as a function of pressure for different 
degrees of polymerization $N$. This behavior reflects the 
well known fact that conformational transitions turn 
into true phase transitions only in the limit 
$N\rightarrow \infty$ \cite{Moore,Grosberg_Khohlov}.

Finally, we estimate the range of  thermodynamic parameters in physical units
at which the local solvent evaporation near the polymer chain can occur.

The  dimensionless parameters are chosen as in fig.3 (a,b): 
$\tilde{T}=0.27$ and $\tilde{\rho}=0.7$ (below the critical isotherm).
In order to obtain an estimate we choose the parameters as $a_{s}=3.64~L^2bar/mol^2$ and 
$v_{s}=0.04~L/mol$ and $a_{s}=5.54~L^2bar/mol^2$ and $v_{s}=0.03~L/mol$, which correspond 
to carbon dioxide ($CO_{2}$) and water, respectively \cite{WanderVaals_fluid}. 
For $CO_{2}$ we obtain $T\approx 280~K$, $P\approx 280~bar$ 
and for water  $T\approx 590~K$, $P\approx 830~bar$. 
These estimates show that the discussed local solvent evaporation 
may be observed under experimentally accessible conditions.

\section{Summary}
Based on a self-consistent field theory taking the solvent explicitly into account, 
we have described two new effects: a polymer collapse due to the strong polymer-solvent 
repulsion accompanied by the solvent evaporation within gyration volume and a globule-coil 
transition at high solvent pressures in the bulk solution accompanied by the solvent 
condensation near the polymer backbone.

Thus, the theory provides at the mean-field level a quantitative explanation of 
the dewetting-induced polymer chain collapse, predicted by ten Wolde and Chandler 
\cite{Chandler_2002}.

As a possible extension of the presented theory the solvent concentration fluctuations 
could be taken into account. This would lead to an additional correction term
in the expression of the total free energy which is related to the 
so-called short-ranged solvent-mediated interactions as described by Fisher and de Gennes \cite{Fisher_1978}. 
Accounting for these non direct fluctuation interactions leads to a renormalization of second virial 
coefficient of the monomer-monomer interaction \cite{Vilgis,Erukhimovich_JETP,Budkov_2014}.
Such approach has been devised describing the hydrophobic effect at 
small and large scales by coarse-grained models and incorporating 
solvent density fluctuations  \cite{tenWolde_2001,Willard_2008,Chandler_2011}.
The role of the solvent density fluctuations in the vicinity of the 
critical point in thermodynamics of dilute polymer solutions has been investigated in works
\cite{Erukhimovich_JETP,Vilgis} within a field theoretical approach.   
We believe, that such corrections will not change qualitatively our final mean-field results, 
although it will become significant in the vicinity of the critical point.

Concluding we would like to speculate about possible applications of the presented theory.
We believe that our theoretical model could find use when interpreting experimental data on 
solubility of polymers in supercritical solvents \cite{Solubility_1,Solubility_2,Solubility_3}. 
Secondly, we believe that the present theory could be used to describe the protein unfolding at high pressures, 
which has been observed in experiments \cite{Winter_1}.

This would, however, require an extension of the theory taking into account hydrogen bond 
formation between polymer backbone and the solvent molecules, incorporating 
attractive interaction between a fraction of monomers. Many-body effects arising from 
long range electrostatic interactions would have to be taken into account as well. 
Moreover, the refinements of the theory would allow to address the protein folding problem 
and the structural rearrangements of the protein globule (liquid-solid and solid-solid transitions). 
However, at the level of the mean-field theory, which does not take into account the effect of 
short ranged particle correlations, the liquid-liquid and liquid-solid transitions 
will remain indistinguishable.

\section*{Acknowledgments}
The research leading to these results has received funding from the European
Union's Seventh Framework Program (FP7/2007-2013) under grant
agreement N//247500 with //project acronym "Biosol". Yu.A.B. and M.G.K. 
thank Russian Scientific Foundation (grant N 14-33-00017).

\begin{figure}[h]
\center{\includegraphics[width=0.7\linewidth]{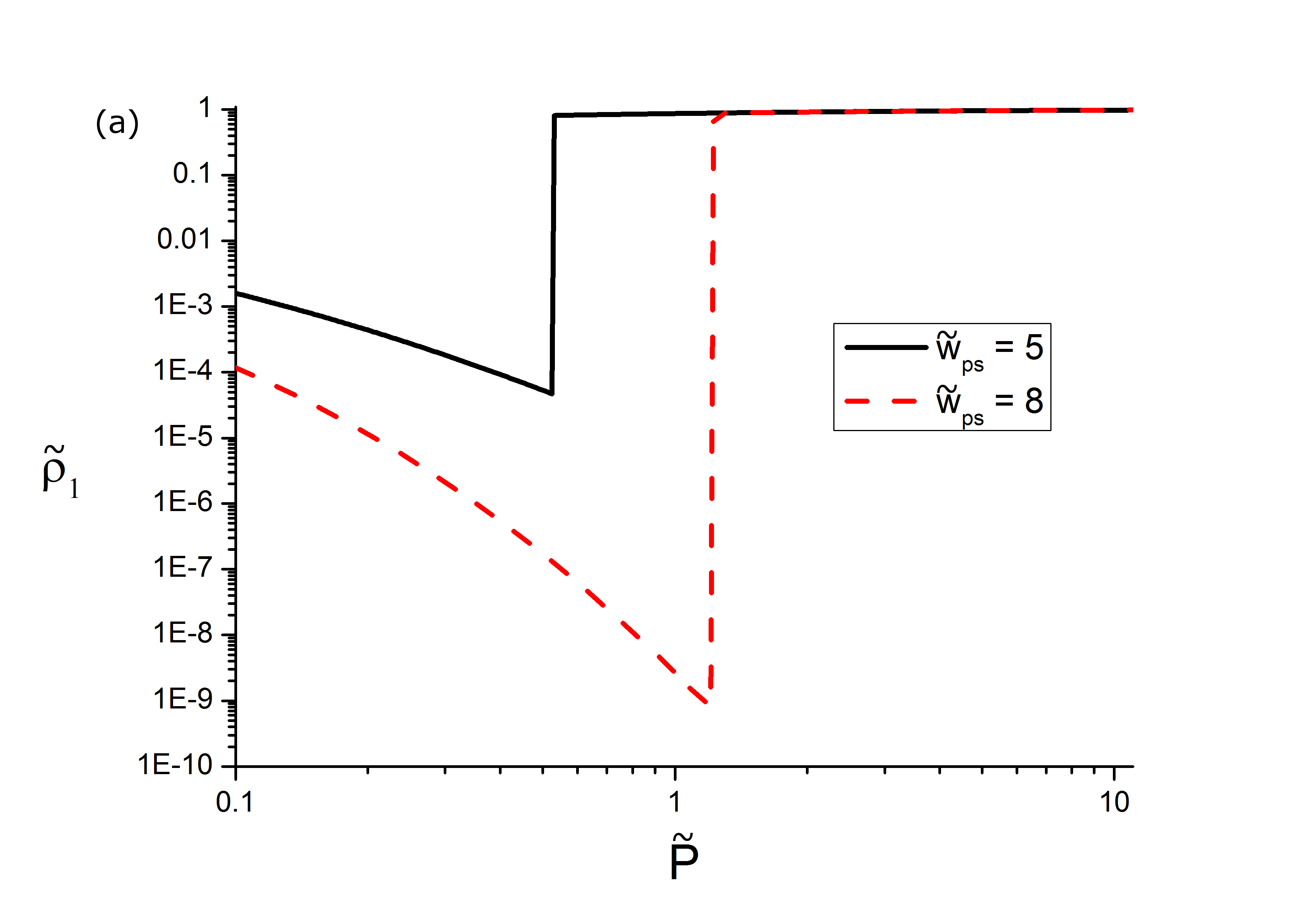}}
\center{\includegraphics[width=0.7\linewidth]{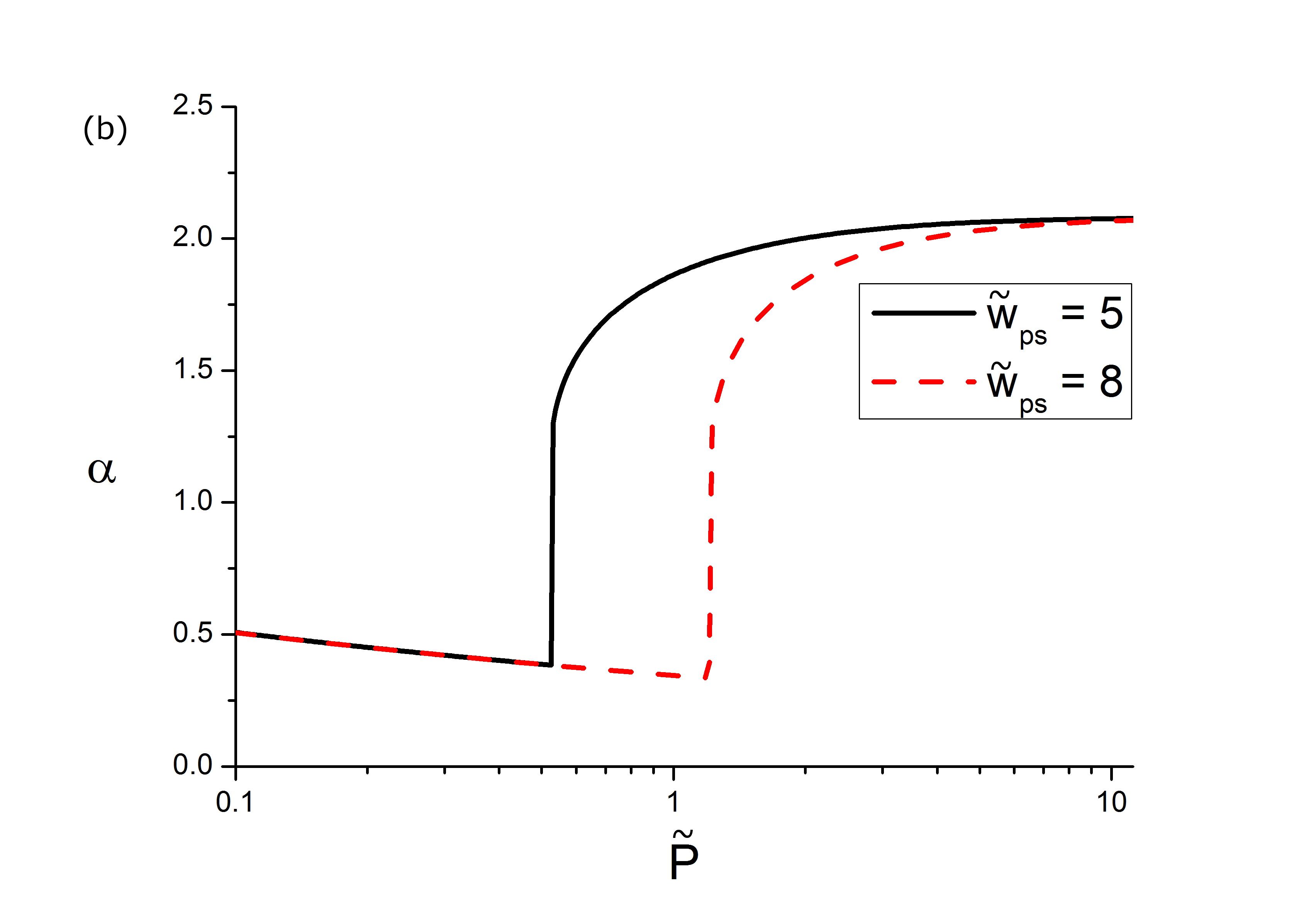}}
\caption{\sl The liquid-phase region of the solvent.
(a) The average solvent concentration in the gyration volume $\tilde{\rho}_{1}$
as a function of the solvent pressure in the bulk $\tilde{P}$ shown for two different solvophobic strength
$\tilde{w}_{ps}=5;8$. (b) The expansion factor $\alpha$ as a function of the solvent
pressure in the bulk solution $\tilde{P}$ shown for the same solvophobic strengths $\tilde{w}_{ps}=5;8$.
Values are shown for $\tilde{v}_{s}=1$, $\tilde{w}_{p}=1$, $N=10^3$, $\tilde{T}=0.27$.}
\label{Fig:1}
\end{figure}

\begin{figure}[h]
\center{\includegraphics[width=0.7\linewidth]{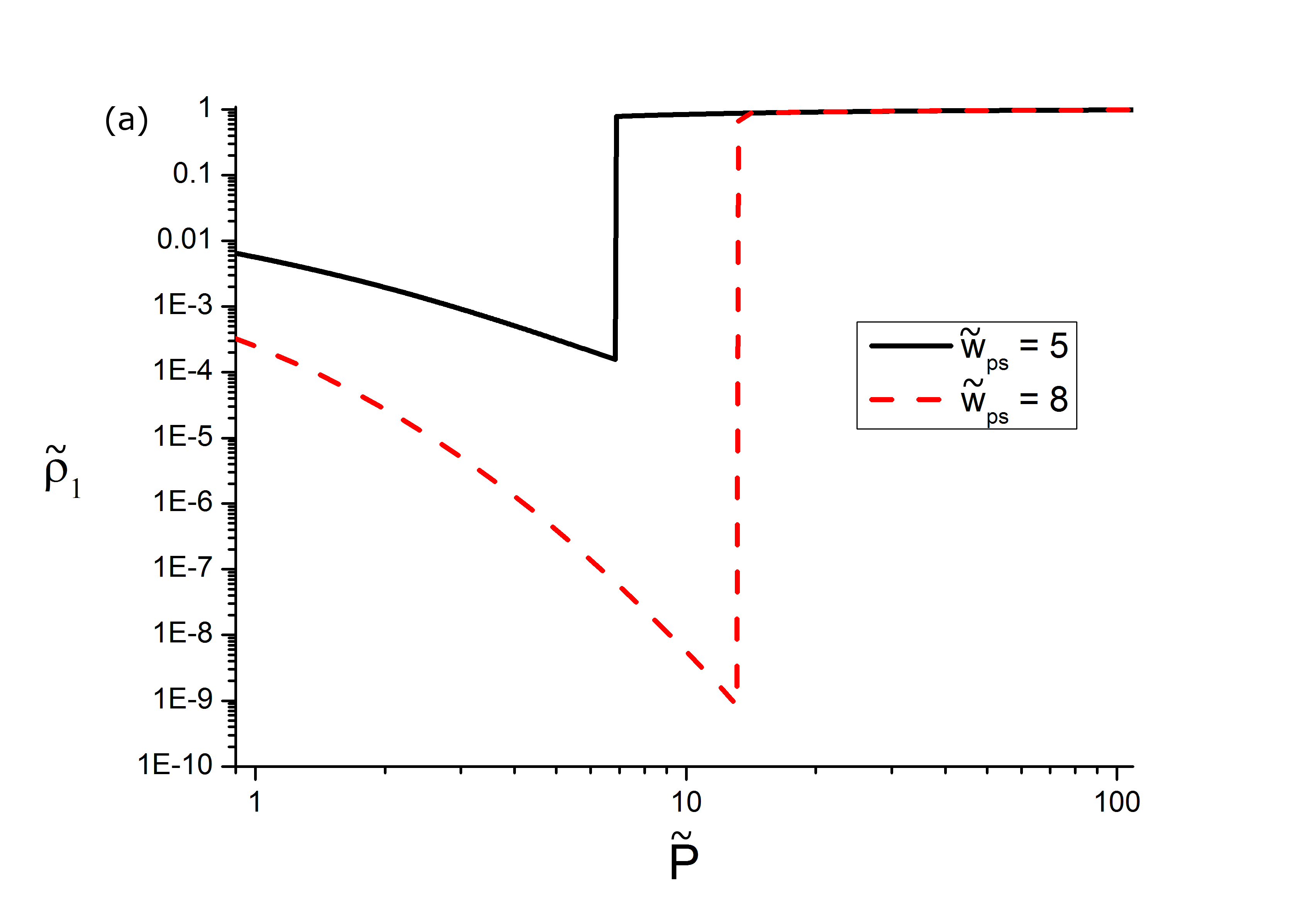}}
\center{\includegraphics[width=0.7\linewidth]{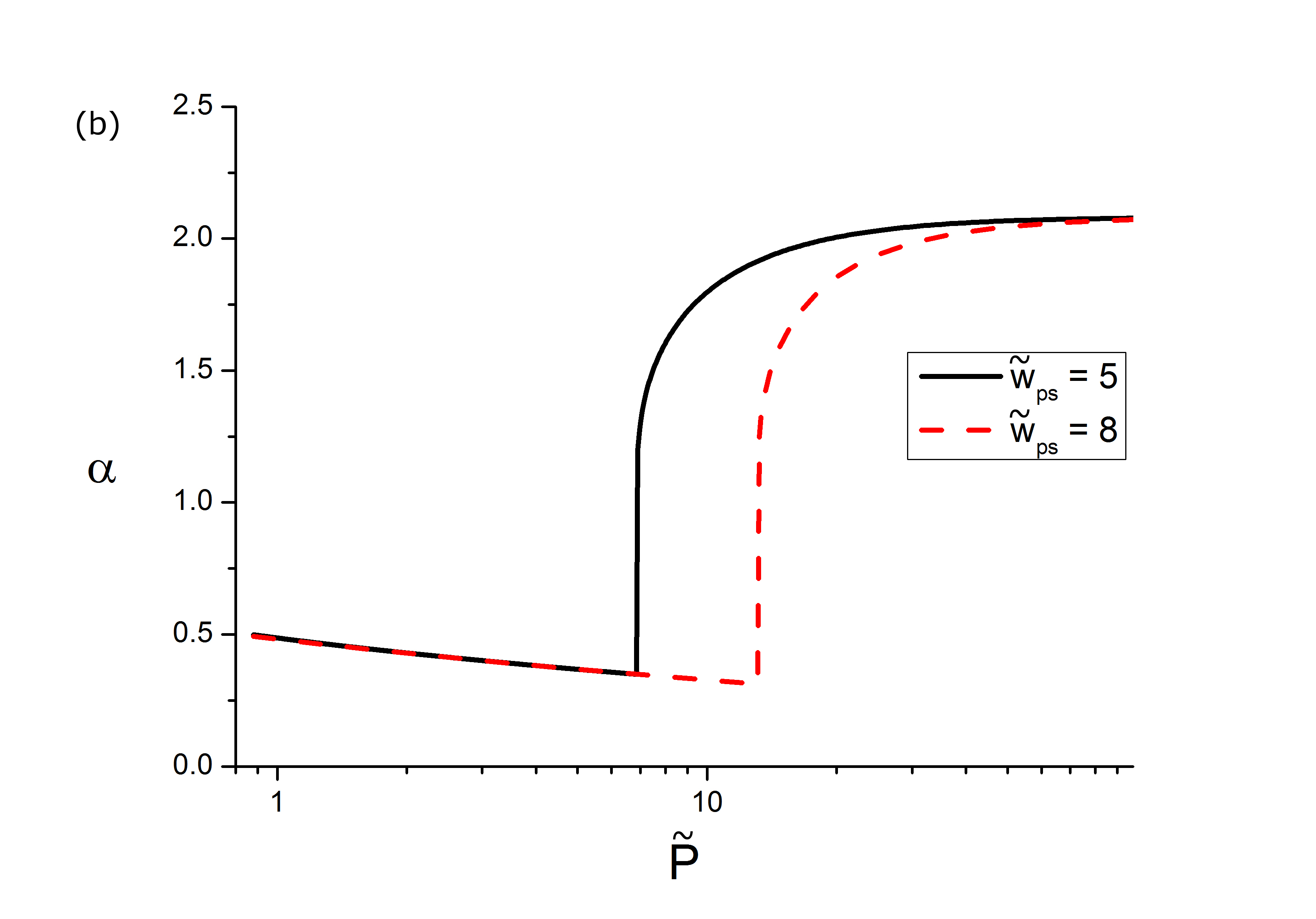}}
\caption{\sl The supercritical region of the solvent ($\tilde{T}>\tilde{T}_{c}$,~ $\tilde{\rho}>\tilde{\rho}_{c}$).
(a) The average solvent concentration in the gyration volume $\tilde{\rho}_{1}$ 
as a function of the solvent pressure in the bulk $\tilde{P}$ 
shown for the solvophobic strength  $\tilde{w}_{ps}=5;8$. (b) The expansion factor $\alpha$ as
a function of the solvent pressure in the bulk solution $\tilde{P}$ shown for the same
solvophobic strength $\tilde{w}_{ps}=5;8$. The behavior of the expansion factor
and the average solvent concentration within the gyration volume is quite similar to the behavior
in region below critical isotherm ($\tilde{T}<\tilde{T}_{c}$). 
Values are shown for $\tilde{v}_{s}=1$, $\tilde{w}_{p}=1$, $N=10^3$, $\tilde{T}=2$.}
\label{Fig:2}
\end{figure}

\begin{figure}[h]
\center{\includegraphics[width=0.6\linewidth]{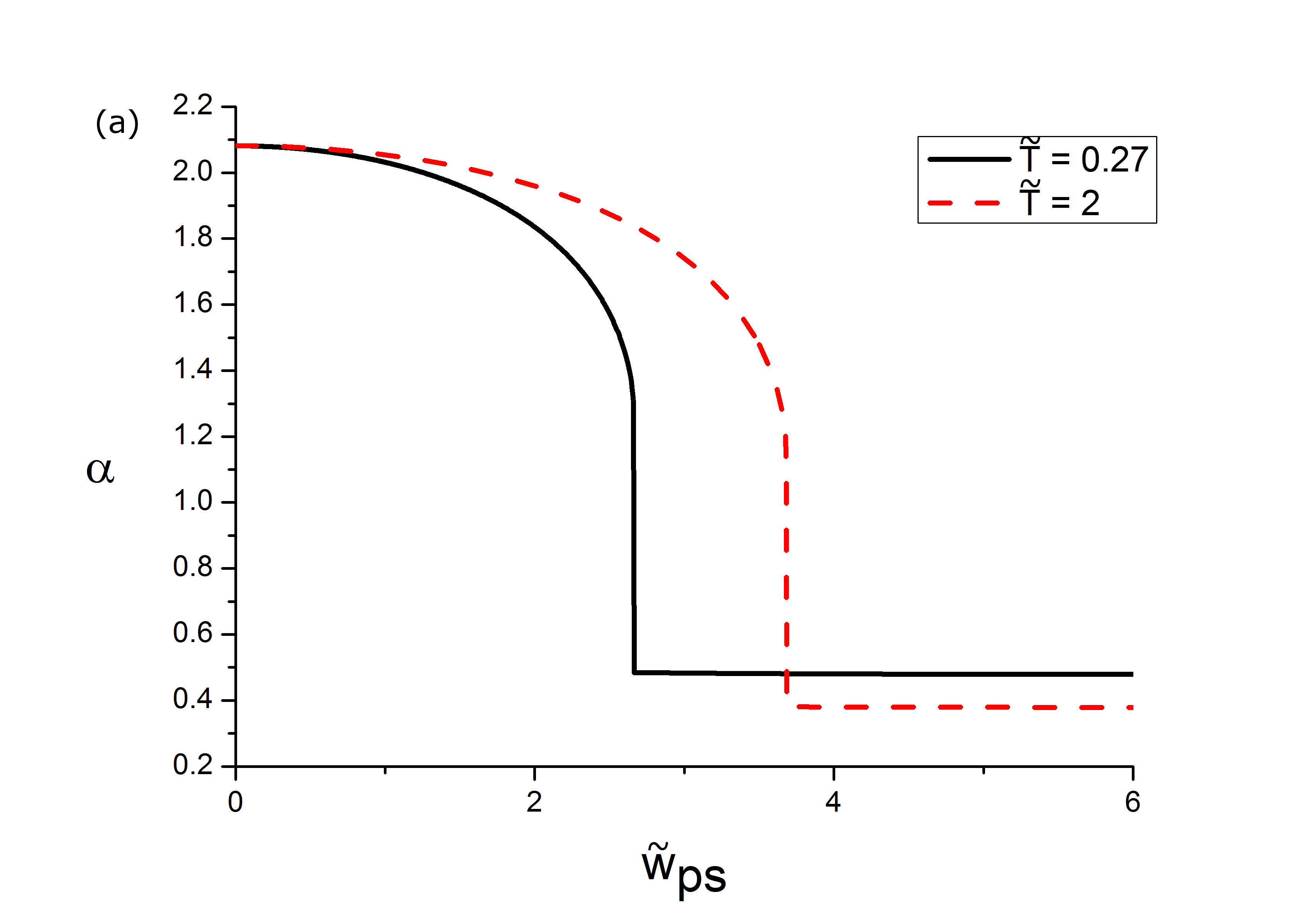}}
\center{\includegraphics[width=0.6\linewidth]{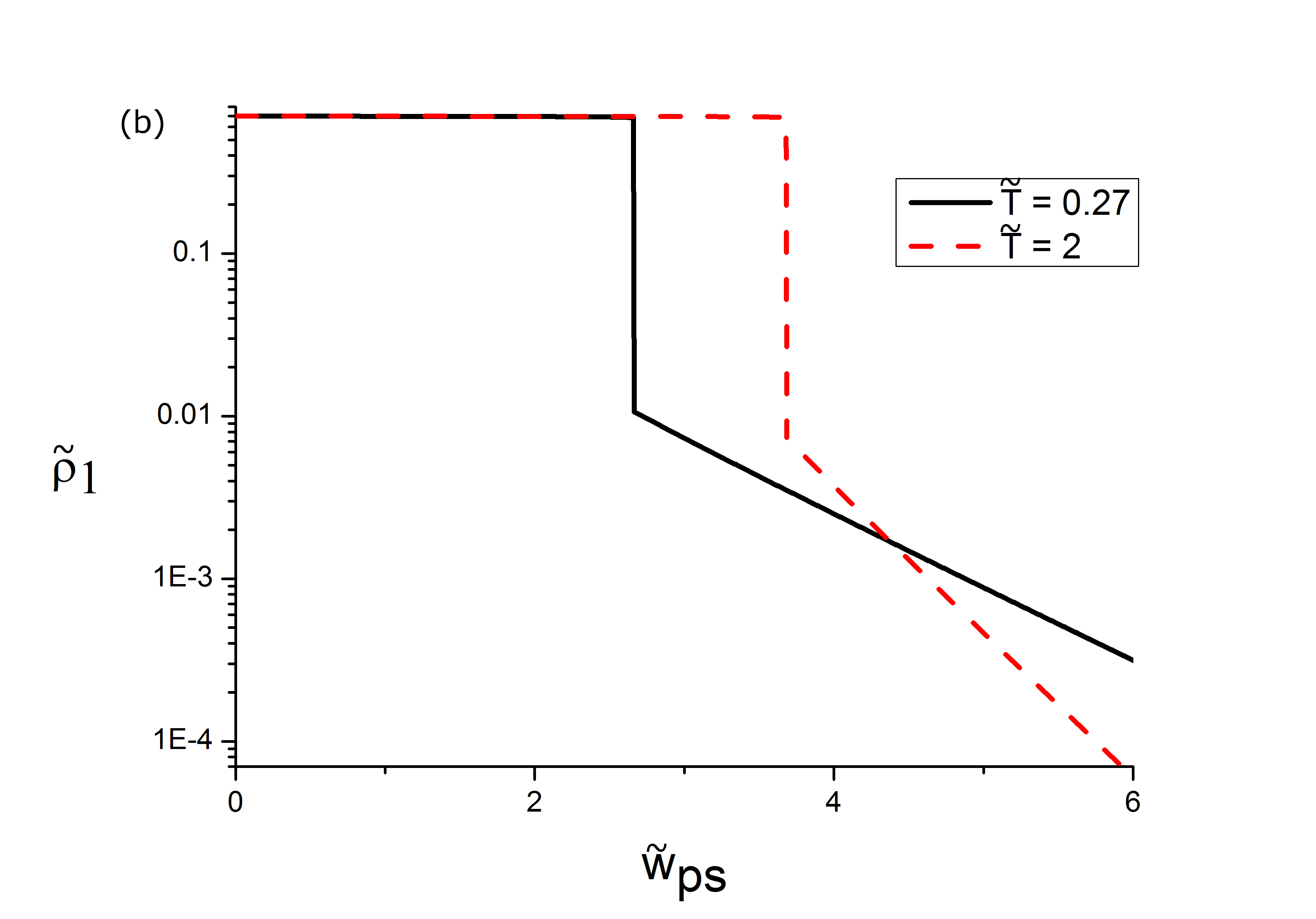}}
\caption{\sl (a) The expansion factor $\alpha$ as a function of 
solvophobic strength $\tilde{w}_{ps}$ at fixed solvent concentration in the bulk  ($\tilde{\rho}=0.7$) 
below and above the critical isotherm of the solvent. (b) 
The average solvent concentration within the gyration volume $\tilde{\rho}_{1}$ as
a function of solvophobic strength $\tilde{w}_{ps}$ at fixed solvent concentration
in the bulk ($\tilde{\rho}=0.7$) below and above the critical isotherm of the solvent. 
In both cases at some threshold value of solvophobic strength the polymer chain 
collapse occurs. For a polymer chain collapse above the critical isotherm a sufficiently higher 
solvophobic strength is required than that in region below critical isotherm. 
Polymer chain collapse occurs as a first-order phase transition.
Values are shown for $\tilde{v}_{s}=1$, $\tilde{w}_{p}=1$, $N=10^3$.}
\label{Fig:3}
\end{figure}

\begin{figure}[h]
\center{\includegraphics[width=0.7\linewidth]{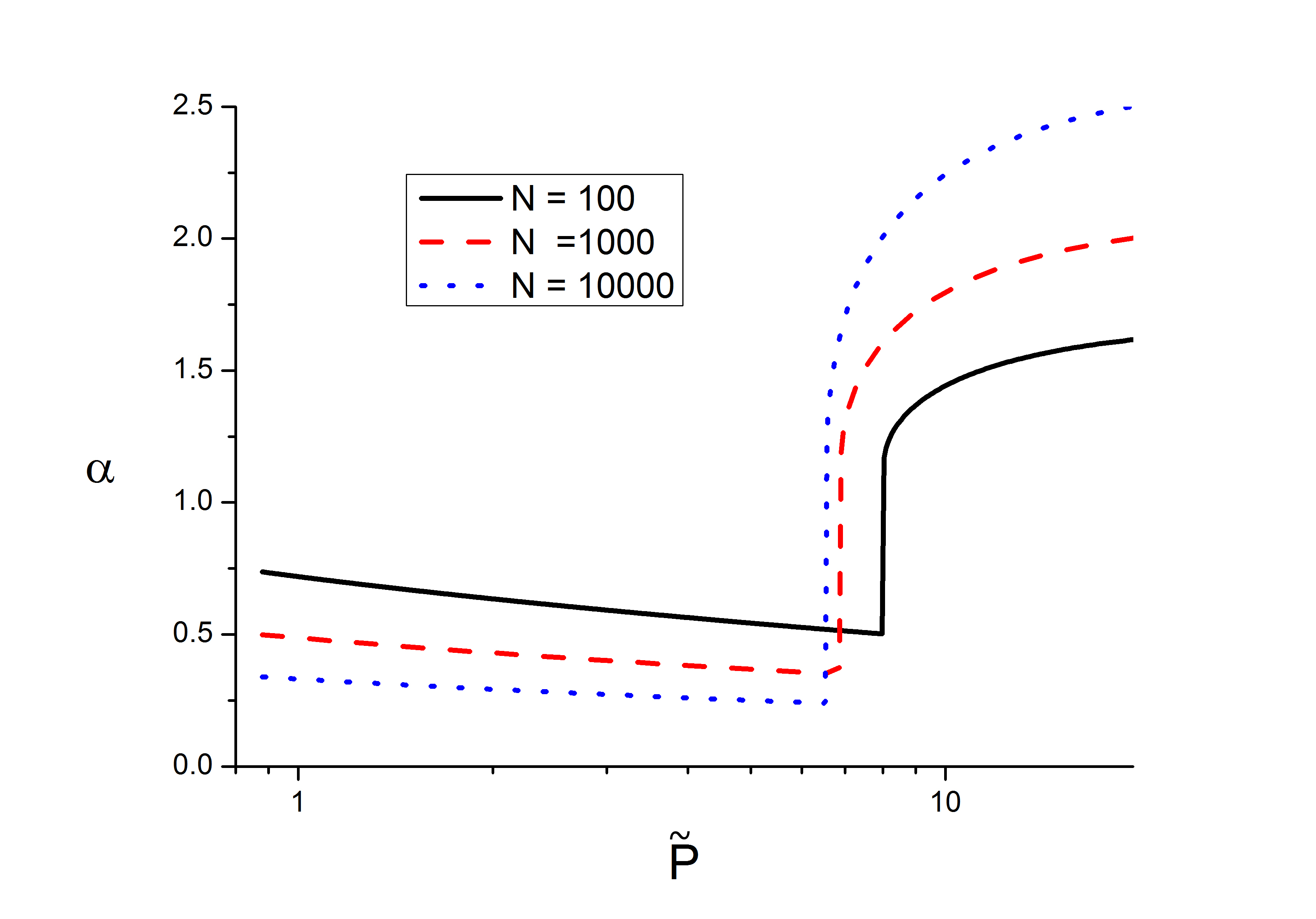}}
\caption{\sl The expansion factor $\alpha$ as a function of 
the solvent pressure in the bulk solution $\tilde{P}$ shown for $N=10^2;10^3;10^4$.
At increasing of degree of polymerization the globule-coil transition occurs more dramatically.
Values are shown for $\tilde{v}_{s}=1$, $\tilde{w}_{p}=1$, $\tilde{w}_{ps}=5$.}
\label{Fig:4}
\end{figure}


\newpage
\begin{thebibliography}{00}

\bibitem{Chandler_2002}
{ten Wolde P.R. and Chandler D.}, PNAS $\bold{99}$, 10, p. 6539, (2002).

\bibitem{DeGennes}
{\sl de Gennes P.-G.}, {\sl Scaling Concepts in Polymer Physics}  Cornell University Press, 1979.

\bibitem{Grosberg_Khohlov}
{A.Yu. Grosberg and A. R. Khokhlov}, {\sl Statistical Physics of Macromolecules}  AIP, New York, 1994.

\bibitem{Edwards_1}
{Edwards S.F.}, Proc. Phys. Soc. $\bold{85}$, p. 613, (1965).

\bibitem{Edwards_2}
{Edwards S.F.}, Proc. Phys. Soc. $\bold{88}$, p. 265,  (1966).

\bibitem{Birshtein_1}
{Birshtein T.M. and Pryamitsyn V.A.}, Macromolecules $\bold{24}$, p. 1554, (1991).

\bibitem{Birshtein_2}
{Birshtein T.M. and Pryamitsyn V.A.}, Polymer Science U.S.S.R. $\bold{29}$, 9, p. 2039, (1987).

\bibitem{Moore}
{Moore M.A.},  J. Phys. A: Math. Gen. $\bold{10}$, 2, p. 305, (1977).

\bibitem{Lifshitz_Grosberg}
{Lifshitz I.M. and Grosberg A.Yu}, Soviet physics JETP. - $\bold{38}$, 6, p. 1198, (1974).

\bibitem{Lifshitz}
{Lifshitz I.M.}, Soviet physics JETP $\bold{28}$, 6, p. L-55, (1975).

\bibitem{deGennes_collaps}
{de Gennes P.G.}, Le Journal De Physique - Letters $\bold{36}$, 2, p. 55 (1975).

\bibitem{Muthukumar}
{Muthukumar M.}, J. Chem. Phys. $\bold{81}$, p. 6272, (1984).

\bibitem{Sanchez}
{Sanchez I.C.}, Macromolecules. $\bold{12}$, 5, p. 980, (1979).

\bibitem{Kholodenko}
{Kholodenko A.L. and Freed K.F.}, J. Phys. A: Math. Gen. $\bold{17}$, p. 2703, (1984).

\bibitem{Grosberg1}
{Grosberg A.Yu. and Kuznetsov D.V.}, Macromolecules  $\bold{25}$, p.  1970, (1992).

\bibitem{Grosberg2}
{Grosberg A.Yu. and Kuznetsov D.V.}, Macromolecules $\bold{25}$, p. 1980, (1992).

\bibitem{Grosberg3}
{Grosberg A.Yu. and Kuznetsov D.V.}, Macromolecules $\bold{25}$, p. 1991, (1992).

\bibitem{Grosberg4}
{Grosberg A.Yu. and Kuznetsov D.V.}, Macromolecules $\bold{25}$, p. 1996, (1992).

\bibitem{Khohlov}
{Khohlov A.R.}, Physica $\bold{105}$, p. 357, (1981).

\bibitem{Khohlov_2}
{J. M. P. van den Oever, F. A. M. Leermakers, G. J. Fleer, V. A. Ivanov, 
N. P. Shusharina, A. R. Khokhlov, and P. G. Khalatur}, Phys. Rev. E. $\bold{65}$, p. 041708, (2002).

\bibitem{Flory_Schultz}
{Schultz R.C. and Flory P.}, J. Polym. Sci. $\bold{15}$, p. 231, (1955).

\bibitem{deGennes2}
{Brochard F. and de Gennes P. G.}, Ferroelectrics. $\bold{30}$, p. L-59, (1980).

\bibitem{Vilgis}
{Vilgis T., Sans A., and Jannink G.}, J. Phys. II France $\bold{3}$, p. 1779, (1993).

\bibitem{Vilgis_Dua}
{Dua A. and Vilgis T.A.}, Macromolecules, $\bold{40}$, p. 6765, (2007).

\bibitem{Dua}
{Dua A. and Cherayil B.J.}, J. Chem. Phys. $\bold{111}$, 7, p. 3274, (1999).

\bibitem{Lifshitz_Grosberg}
{Lifshitz I.M., Grosberg A.Yu., A.R. Khohlov}, Rev. Mod. Phys. $\bold{50}$, 3, p. 683, (1978).

\bibitem{Erukhimovich_JETP}
{Erukhimovich I.Ya.}, Journal of Experimental and Theoretical Physics $\bold{87}$, 3, p. 494, (1998).

\bibitem{Tanaka}
{Matsuyama M. and Tanaka F.}, J. Chem. Phys., $\bold{94}$, 1, p. 781, (1991).

\bibitem{Tanaka2}
{Tanaka F., Koga T. and Winnik F.M.}, Phys. Rev. Lett., $\bold{101}$, p. 028302, (2008).

\bibitem{Budkov_2014}
{Budkov Yu.A., Kolesnikov A.L., Georgi N., and Kiselev M.G.}, J. Chem. Phys. $\bold{141}$, p. 014902, (2014).

\bibitem{Frederickson_book}
{Fredrickson G. H.} {\sl The Equilibrium Theory of Inhomogeneous Polymers} Oxford: Clarendon Press, 2006.

\bibitem{Fixman_Gyration}
{Fixman M.}, J. Chem. Phys. $\bold{36}$, 2, p. 306, (1962).

\bibitem{Flory_book}
{Flory P.}, {\sl Statistical Mechanics of Chain Molecules}  New York: Wiley-Interscience, 1969.

\bibitem{Hansen}
{Barrat J.-L. and Hansen J.-P.} {\sl Basic Concepts for Simple and Complex Liquids} University Press, Cambridge, 2003.

\bibitem{tenWolde_2001}
{ten Wolde P.R., Sun S.X., and Chandler D.}, Phys. Rev. E. $\bold{65}$, p. 011201, (2001).

\bibitem{Willard_2008}
{Willard A.P. and Chandler D.}, J. Phys. Chem. B $\bold{112}$, p. 6187, (2008).

\bibitem{Chandler_2011}
{Varilly P., Patel A.J., and Chandler D.}, J. Chem. Phys. $\bold{114}$, p. 074109, (2011).

\bibitem{Fisher_1978}
{Fisher, M. E. and de Gennes, P. G.}, C. R. Acad. Sci. Paris B. $\bold{287}$, p. 207, (1978).

\bibitem{WanderVaals_fluid}
{Weast R. C.} {\sl Handbook of Chemistry and Physics} Cleveland: Chemical Rubber Co., 1972.

\bibitem{Solubility_1}
{Dardin A., Cain J.B., DeSimone J.M., Johnson, Jr C.S., Samulski E.T. }, Macromolecules $\bold{30}$, 12, p. 3593, (1997).

\bibitem{Solubility_2}
{Andre P., Lacroix-Desmazes P., Taylor D.K., and Boutevin B.}, J. of Supercritical Fluids $\bold{37}$, p. 263, (2006).

\bibitem{Solubility_3}
{Kiran E.}, J. of Supercritical Fluids $\bold{47}$, p. 466, (2009).

\bibitem{Winter_1}
{Herberhold H. and Winter R.} // Biochemistry $\bold{41}$, p. 2396, (2002).

\bibitem{Kubo}
{Kubo R.}, J. Phys. Soc. Jap. $\bold{17}$, p. 1100, (1962).

\bibitem{Tamm}
{Tamm M.V., Erukhimovich I.Ya.}, Polymer Science, Ser. A, $\bold{44}$, 2, p. 196, (2002).



\end{thebibliography}
\end{document}